\documentclass[twocolumn,showpacs,preprintnumbers,amsmath,amssymb]{revtex4}

\usepackage{graphicx}
\usepackage{dcolumn}
\usepackage{bm}

\begin{document}

\preprint{APS/123-QED}

\title{Evaluations of pressure-transmitting media for cryogenic experiments \\with diamond anvil cell\footnote{Review of Scientific Instruments {\bf 80}, 123901 (2009)}}

\author{Naoyuki Tateiwa}
 \email{tateiwa.naoyuki@jaea.go.jp} 
\author{Yoshinori Haga}%

\affiliation{Advanced Science Research Center, Japan Atomic Energy Agency, Tokai, Naka, Ibaraki 319-1195, Japan}%

\date{\today}

\begin{abstract}
The fourteen kinds of pressure-transmitting media were evaluated by the ruby fluorescence method at room temperature, 77 K using the diamond anvil cell (DAC) up to 10 GPa in order to find appropriate media for use in low temperature physics. The investigated media are a 1:1 mixture by volume of Fluorinert FC-70 and FC-77, Daphne 7373 and 7474, NaCl, silicon oil (polydimethylsiloxane), Vaseline, 2-propanol, glycerin, a 1:1 mixture by volume of $n$-pentane and isopentane, a 4:1 mixture by volume of methanol and ethanol, petroleum ether, nitrogen, argon and helium. The nonhydrostaticity of the pressure is discussed from the viewpoint of the broadening effect of the ruby $R_1$ fluorescence line. The $R_1$ line basically broadens above the liquid-solid transition pressure at room temperature. However, the nonhydrostatic effects do constantly develop in all the media from the low-pressure region at low temperature. The relative strength of the nonhydrostatic effects in the media at the low temperature region is discussed. The broadening effect of the ruby $R_1$ line in the nitrogen, argon and helium media are significantly small at 77 K, suggesting that the media are more appropriate for cryogenic experiments under high pressure up to 10 GPa with the DAC. The availability of the three media was also confirmed at 4.2 K. 
\end{abstract}

\maketitle

\section{Introduction}
The development of the diamond anvil cell (DAC) has revolutionized high-pressure research in various scientific fields such as physics, chemistry, geophysics, materials science and biology~\cite{jayaraman1,eremets}. Various types of experiments can be performed with the DAC. For instance,  X-ray diffraction experiments, optical absorption and reflectivity measurements, Raman and Brillouin scattering studies, and transport and thermal measurements under high pressure above 10 GPa have become possible. These methods have been successfully applied in low temperature physics where many interesting pressure-induced phenomena such as superconductivity have been extensively studied~\cite{buzea,shimizu1}. 
  
 In high-pressure experiments, nonhydrostatic effects such as inhomogeneous pressure distribution (pressure gradient) and uniaxial (deviatoric) stress need to be reduced because the effects can strongly influence the physical state of a sample. It is important to use a pressure-tranmistting medium in a pressure region where it is in a liquid state. However, a liquid medium solidifies through a liquid-solid transition on cooling from room temperature.  The shear stress of a solid medium causes nonhydrostatic pressure in the low temperature region.
  
 The purpose of the present study is to evaluate the pressure-qualities of media used in low temperature physics, and in particular for the strongly correlated electron system in which novel types of physical phenomena such as non-Fermi liquid behavior or unconventional superconductivity have been extensively studied~\cite{flouquet,onuki}.  The electronic state of the system is generally sensitive to small amounts of impurities or nonhydrostatic effects. A recent example is the case of the ``FeAs-based superconductor" CaFe$_2$As$_2$ in which superconductivity was discovered above 0.5 GPa in high-pressure studies using organic media~\cite{torikachvili,park}, while it was not observed with a helium medium~\cite{yu}. This discrepancy may result from sensitive structural instability to the uniaxial stress of the compound. nonhydrostatic effects cannot be avoided in the low temperature region. It is necessary to compare quantitatively the strength of the nonhydrostatic effects of various kinds of pressure-media. 
 
  Most of previous studies on pressure-media have concerned the hydrostaticity of the pressure at room temperature~\cite{piermarini1,angel,osakabe,klotz}. Very few studies have been carried out on the pressure-qualities of media in the low temperature region up to 3 GPa~\cite{burnett,fukazawa}. In this study, therefore, the fourteen kinds of pressure-transmitting media were studied by the ruby fluorescence technique up to 10 GPa at room temperature, 77 and 4.2 K.      
    
\section{Experimental}
 The fourteen kinds of investigated media are a 1:1 mixture by volume of Flourinert FC-70 and FC-77 (Flourinert FC70/77, Sumitomo 3M), Daphne 7373 and 7474 (Idemitsu Kosan), NaCl, Vaselin, silicon oil (polydimethylsiloxane with a kinematic viscosity of 1 mm$^{2}{\cdot}$s$^{-1}$, Shin-Etsu Chemical), 2-propanol, glycerin, 1:1 mixture by volume of $n$-pentane and isopentane (pentane mixture), a 4:1 mixture by volume of methanol and ethanol (4:1 M-E mixture), petroleum ether (Wako Chemical Industries), nitrogen, argon and helium. The strength of the nonhydrostatic effects in the media was evaluated by the ruby (Al$_2$O$_3$: Cr$^{3+}$) fluorescence method using a clamp-type diamond anvil cell originally designed by Dunstan and Spain~\cite{syassen,dunstan1,dunstan2}. Figure 1 shows a schematic illustration of the DAC setup and the fluorescence spectra of the ruby $R$ lines for the representative media of NaCl, glycerin, 4:1 M-E mixture and argon at 77 K and around 10 GPa. The culet of the diamond was 800 ${\mu}$m. A sample chamber of 400 ${\mu}$m in diameter was prepared in a 304 stainless steel gasket.  Small ruby chips, containing 0.5 wt\% of Cr$_2$O$_3$, were uniformly placed in the chamber filled with a pressure-transmitting medium. The diameter of the chips was less than 10 $\mu$m.  It was confirmed by X-ray diffraction analysis that the chips were single crystals.

 The ruby fluorescence was measured using a charged coupled device (CCD) spectrometer. More details on the optical system are given in Reference 19.  A diode pumped solid-state (DPSS) green laser light (532 nm), introduced into the sample chamber through fiber optics, excited the ruby chips in the chamber. The diameter of the beam line was 600 $\mu$m, larger than that of the sample chamber.  The present work reveals spatially averaged information on the nonhydrostatic effects in the whole sample chamber.
  
  When the pressure becomes nonhydrostatic, the two $R_1$ and $R_2$ lines broaden. The pressure shift of the ruby $R_1$ line is sensitive to the uniaxial stress~\cite{chai,adams,he}. In the present study, a number of small ruby single crystals were uniformly placed inside the chamber with the directions of the single crystals being random. The broadening of the ruby $R_1$ line reflects both the inhomogeneous pressure distribution and uniaxial stress pressure. 

  The peak positions and line widths of the ruby $R_1$ and $R_2$ fluorescence lines were determined by deconvoluting measured $R$-line spectra into a pair of pseudo-Voigt functions that represent the contribution of the ruby $R_1$ and $R_2$ lines with a linear background. The pressure was determined using the hydrostatic ruby pressure scale by Zha, Mao and Hemley~\cite{zha,kawamura}.  
    \begin{figure}
\includegraphics[width=7.5cm]{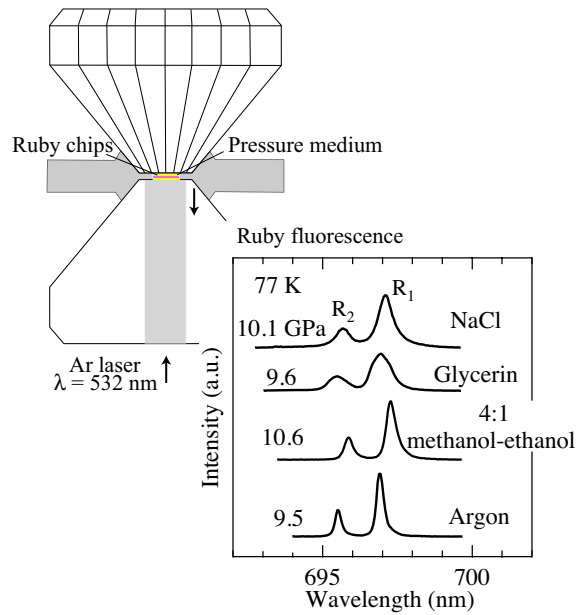}
\caption{\label{fig:epsart}(Color online) Schematic illustration of the DAC set up and the fluorescence spectra of the ruby $R$ lines with representative media of NaCl, glycerin, 4:1 M-E mixture and argon at 77 K and around 10 GPa}
\end{figure}
  
  The pressure-transmitting medium was loaded into the sample chamber at room temperature except for nitrogen, argon and helium done using a purpose built cryogenic device at 77 or 1.4 K~\cite{dunstan1,dunstan2}. The pressure was applied and changed at room temperature. The ruby fluorescence spectra were measured after the pressure and the width of the ruby $R_1$ line had been stabilized. It took several hours for the quantities to be stabilized when a medium was in a solid state under high-pressure. The DAC was slowly cooled down to 77 K and 4.2 K using liquid nitrogen and helium, respectively. It should be noted here that the pressure did not significantly change during the cooling process with the present DAC where load was maintained using springs. The difference in pressure between at 4.2 K and 300 K was in the order of a few $\%$ when organic and argon media were used. The pressure-change with the nitrogen and helium media will be discussed later.
 
   Here we note differences between the present work and the previous studies on pressure-media from a technical point of view. The references 10 and 13 reported the inhomogeneous pressure distribution inside the sample chamber at room temperature~\cite{piermarini1,klotz}. The previous studies in references 11, 12, 14 and 15 revealed spatially local information on the nonhydrostatic effects around the center of the chamber as the size of the ``sensors'' used to detect the effects, for example the ruby chips, Cu$_2$O or NaCl, was very small compared with the sample space~\cite{angel,osakabe,burnett,fukazawa}. This work shows spatially averaged information on the nonhydrostatic effects inside the sample chamber. The observed broadening effect of the ruby $R_1$ line reflects both the inhomogeneous pressure distribution and uniaxial stress pressure. The purpose of the present study is to clarify the relative strength of the nonhydrostatic effects of the media in the low temperature region.
      
     As shown in Figure 1, the broadening effect on the $R_1$ line depends on the media at 77 K and around 10 GPa. A clear broadening effect was observed in the ruby $R_1$ lines for NaCl and glycerin. However, the lines for the 4:1 M-E mixture and argon are comparably sharper. Therefore, it is possible to discuss the relative strength of the nonhydrostatic effects of the media from the width of the ruby $R_1$ line.

  \section{Results}
 In this section, we show the pressure dependence of the full-width at half maximum (FWHM) of the ruby $R_1$ line with the media.  The pressure dependence was studied several times with independent settings for each of the media. The guide lines in the figures are fitted curves of the pressure dependences of the FWHMs with polynomial functions. The open circles and squares indicate the FWHMs at room temperature in the first and second runs, respectively. The open up and down-pointing triangles indicate the FWHMs at 77 K in the first and second runs, respectively. The solid triangles indicate the FWHMs at 4.2 K. The nonhydrostatic effects are discussed from the value of $\Delta$FWHM($P$) (= FWHM($P$)-FWHM(0)). Here, FWHM($P$) and FWHM(0) are the line widths of the ruby $R_1$ line at high pressure and ambient pressure, respectively, obtained by the analyses of the ruby profile with the Voigt function where the instrumental resolution was taken into account. Depending on the size of the additional broadening $\Delta$FWHM at 5 GPa and 77 K, the media are classified into three groups (I, II and III).
 
 \subsection{\label{sec:level2}Group I: 1:1 mixture of Fluorinert FC-70 and FC-77, Daphne 7373, NaCl, silicon oil (polydimethylsiloxane) and Vaseline}

  Fluorinerts and Daphne 7373 have been widely used in high pressure studies using piston cylinder type high-pressure cells because the media have the advantage of being chemically inert~\cite{sidorov1,murata1,murata2}. The hydrostatic-limit pressures (solidification pressures) of Flourinert FC70/77 and Daphne 7373 at room temperature were determined to be 1.2 and 2.4 GPa, respectively. 
    
  Figures 2 (a) and (b) show the pressure dependences of the FWHMs for Fluorinert FC70/77 and Daphne 7373.  At room temperature, the FWHMs start to increase above the solidification pressures. The value of FWHMs for the media is about 1.1 nm at 10 GPa, indicating strong nonhydrostatic effects.  A strong broadening effect was also observed at 77 K.  The value of $\Delta$FWHMs for the media is about 0.7 nm at 10 GPa. The present results suggest that they are not suitable media in the higher-pressure region.   
    \begin{figure}
\includegraphics[width=7cm]{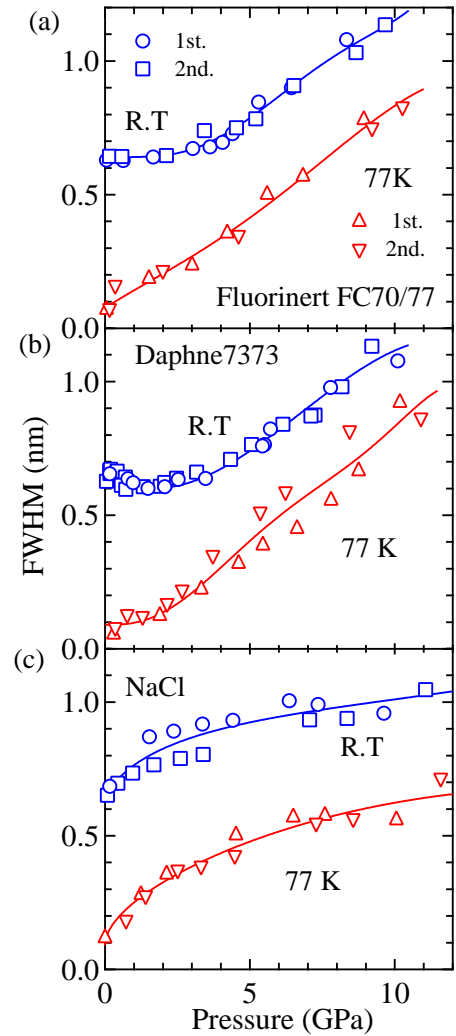}
\caption{\label{fig:epsart} (Color online)Pressure dependence of full-width at half maximum (FWHM) of the ruby $R_1$ fluorescence line at room temperature and 77 K for (a)Fluorinert FC70/77, (b) Daphne 7373, and (c) NaCl. Guide lines in the figures are fitted curves of the pressure dependences of the FWHMs with polynomial functions. }
\end{figure}
       \begin{figure}
\includegraphics[width=7.cm]{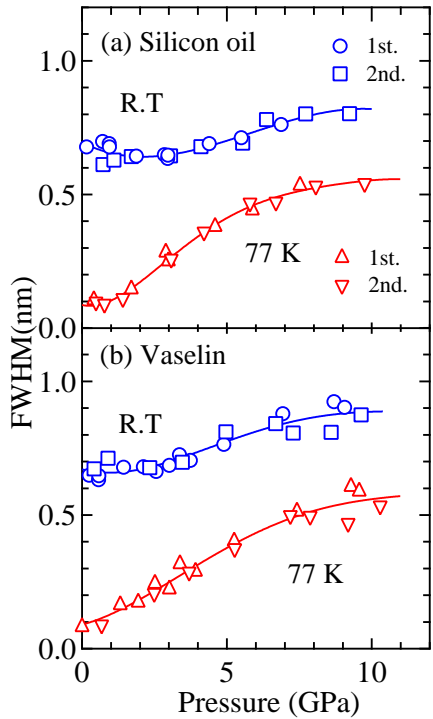}
\caption{\label{fig:epsart}(Color online)Pressure dependence of full-width at half maximum (FWHM) of the ruby $R_1$ luminescence line at room temperature and 77 K for (a)silicon oil (polydimethylsiloxane) and (b) Vaseline.}
\end{figure}

    Figure 2 (c) shows the pressure dependences of the FWHMs for the solid medium of NaCl at room temperature and 77 K. The FWHMs at both temperatures increase non-linearly in the lower pressure region below 1 GPa and show a tendency to saturate at the high pressure region above 5 GPa. The pressure dependence is quite different from those of liquid media. The value of $\Delta$FWHM at 10 GPa is about 0.5 nm at 77 K. It is suggested that the hydrostatic pressure is realized in a very small pressure region even at room temperature.

  The results of the NaCl medium are compared with those of previous studies that reported the pressure distribution inside a sample chamber filled with the NaCl medium becomes spatially inhomogeneous above 4 GPa at room temperature~\cite{piermarini1}.  It should be noted, however, that the observed broadening effect in the present study reflects both the uniaxial stress and inhomogeneous pressure distribution inside the chamber, as mentioned in the previous section. This suggests that uniaxial stress is the main cause of the broadening effect in the low-pressure region.  
    
  We show the experimental results with the silicon oil and Vaseline.  Silicon oils such as polydimethylsilioxane, polyethylsiloxane and hexamethyldisiloxane have been accepted as good media for a long time~\cite{bridgman1,kirichenko,sandberg}. Polydimethylsilioxane ((CH$_3$)$_3$SiO-[SiO(CH$_3$)$_2$]$_n$-(CH$_3$)$_3$Si) with a kinematic viscosity of 1 mm$^{2}{\cdot}$s$^{-1}$ was tested. Vaseline (petroleum jelly or petrolatum) has been occasionally used in high-pressure experiments~\cite{wittig1,sacchetti}. However, there have been no reports on its suitability as a pressure-medium as far as is known.  

    Figure 3 (a) and (b) show the pressure dependences of the FWHMs with silicon oil and Vaselin at room temperature and 77 K. The FWHMs start to weakly increase above 3 GPa. The ruby $R_1$ lines for the media do not show a strong broadening effect at 10 GPa, thus indicating weaker nonhydrostatic effects of the media compared with those of the former media. Previous studies have pointed out that silicon oil is as good a pressure medium as the 4:1 mixture of methanol and ethanol up to 20 GPa at room temperature~\cite{ragan,shen1}. However, a small but finite broadening effect does appear above 3 GPa, thus suggesting the pressure-quality of the silicon oil to be lower than that of the 4:1 methanol-ethanol mixture.  At 77 K, the FWHMs for the media increase rapidly with increasing pressure above about 1 GPa. The value of the $\Delta$FWHMs for the media at 10 GPa is about 0.5 nm, comparable with that of NaCl.

   \subsection{\label{sec:level2}Group II: 2-propanol, glycerin and Daphne 7474}
      
  The secondary alcohol of 2-propanol (Isopropanol: CH$_3$CH(OH)CH$_3$) has been accepted as a good pressure-transmitting medium in high pressure studies. The hydrostatic-limit pressure of 2-propanol was determined to be 4.2 GPa by the ruby fluorescence method at room temperature~\cite{piermarini1}. The trivalent alcohol of glycerin (glycerol, C$_3$H$_5$(OH)$_3$) has been used mainly for medical purposes but also in the high-pressure studies for a long time~\cite{bridgman0,kurita1}. Osakabe and Kakurai pointed out that glycerin is the most suitable pressure-medium for use in neutron scattering experiments under high pressure up to 7 GPa~\cite{osakabe}. There have been no reports on the hydrostatic-limit pressure of the glycerin medium. Daphne 7474 is a transparent and chemically non-reactive medium recently developed by Murata {\it et al.}~\cite{murata3}. The solidification pressure of Daphne 7474 was reported to be 3.7 GPa at room temperature. 
  \begin{figure}[b]
\includegraphics[width=7cm]{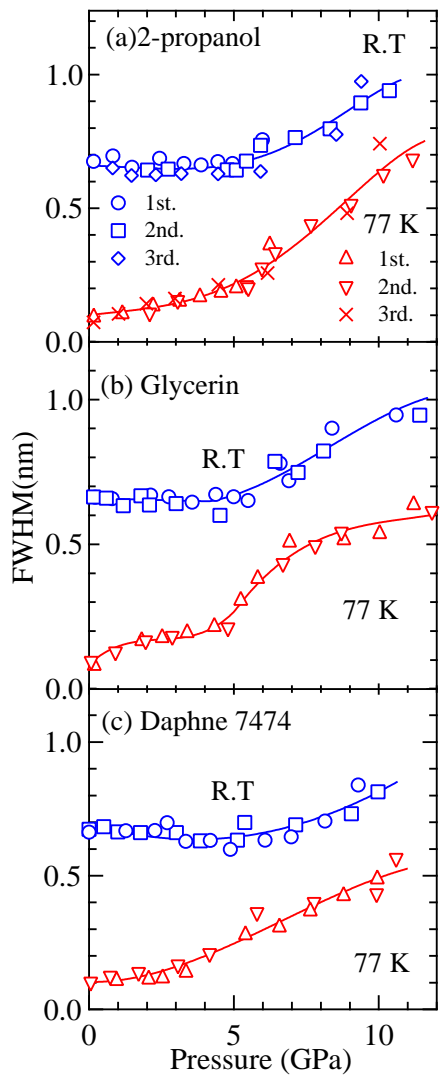}
\caption{\label{fig:epsart}(Color online)Pressure dependence of full-width at half maximum (FWHM) of the ruby $R_1$ luminescence line at room temperature and 77 K for (a) 2-propanol, (b) glycerin and (c) Daphne 7474.}
\end{figure} 
  
 Figures 4 (a), (b) and (c) show the pressure dependences of the FWHMs for 2-propanol, glycerin and Daphne 7474 at room temperature and 77 K. At room temperature, the FWHMs for 2-propanol and Daphne 7474 start to increase roughly above the solidification pressures~\cite{piermarini1,murata3}. The FWHM for the glycerin medium starts to increase above 5 GPa, which roughly corresponds to the solidification pressure at room temperature~\cite{drozd}. This suggests the hydrostatic-limit pressure to be about 5 GPa at room temperature. At 77 K, the FWHMs of the three types of media show different pressure dependences. The FWHMs for 2-propanol and Daphne 7474 weakly increase in the low-pressure region below 5 GPa but then strongly increase above the pressure. The pressure dependence of the FWHM for the glycerin medium shows a complex behavior: it increases non-linearly below 1 GPa, shows a tendency to saturate above 2 GPa, and then strongly increases again with increasing pressure above 5 GPa. The values of the $\Delta$FWHMs for the three media are less than 0.2 nm at 5.0 GPa, significantly smaller than those of the media in Group I. However, the values of the $\Delta$FWHMs at 10 GPa are comparable with those of the media in Group I. 

  The value of the FWHM with glycerin is larger than those of some of the media in Group I, for example, Daphne 7373 and silicon oil, below 2 GPa, thus suggesting the strength of the nonhydrostatic effects with the glycerin medium to be larger than those of the two media. This is consistent with a previous evaluation of pressure-media by the $^{63}$Cu-NQR spectra of Cu$_2$O at 4.2 K~\cite{fukazawa}.

\subsection{\label{sec:level2}Group III: 1:1 mixture of $n$-pentane and isopentan, 4:1 mixture of methanol and ethanol, petroleum ether, nitrogen, argon and helium.}

We firstly show the results of the 1:1 mixture by volume of $n$-pentane and isopentane (pentane mixture), the 4:1 mixture by volume of methanol and ethanol (4:1 M-E mixture) and the petroleum ether. The pentane and 4:1 M-E mixtures have been widely used in high-pressure studies for a long period of time~\cite{bridgman2,piermarini1}. The hydrostatic-limit pressures at room temperature were determined by the ruby fluorescence method to be 7.3 and 10.5 GPa for the pentane and 4:1 M-E mixtures, respectively~\cite{piermarini1}. Petroleum ether (benzine or petroleum naphtha) is a group of liquid hydrocarbon mixtures whose main components are isohexan and $n$-pentane.  It has been mainly used as an industrial washing solution but also in high-pressure studies for a long time~\cite{bancroft,sawaoka}. However, there have been no reports on the hydrostatic-limit pressure as far as is known. Figures 5 (a), (b) and (c) show the pressure dependences of the FWMHs of the three organic media at room temperature and 77 K. 
         \begin{figure}
\includegraphics[width=7.5cm]{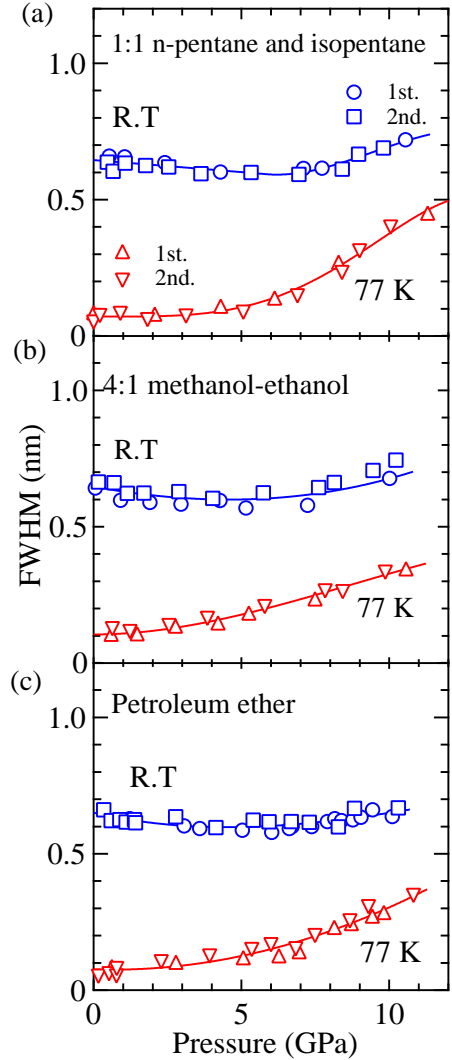}
\caption{\label{fig:epsart}(Color online)Pressure dependence of full-width at half maximum (FWHM) of the ruby $R_1$ luminescence line at room temperature and 77 K for (a)1:1 $n$-pentane and isopentane, (b) 4:1 methanol-ethanol, and (c) petroleum ether.}
\end{figure}
      \begin{figure*}
\includegraphics[width=16cm]{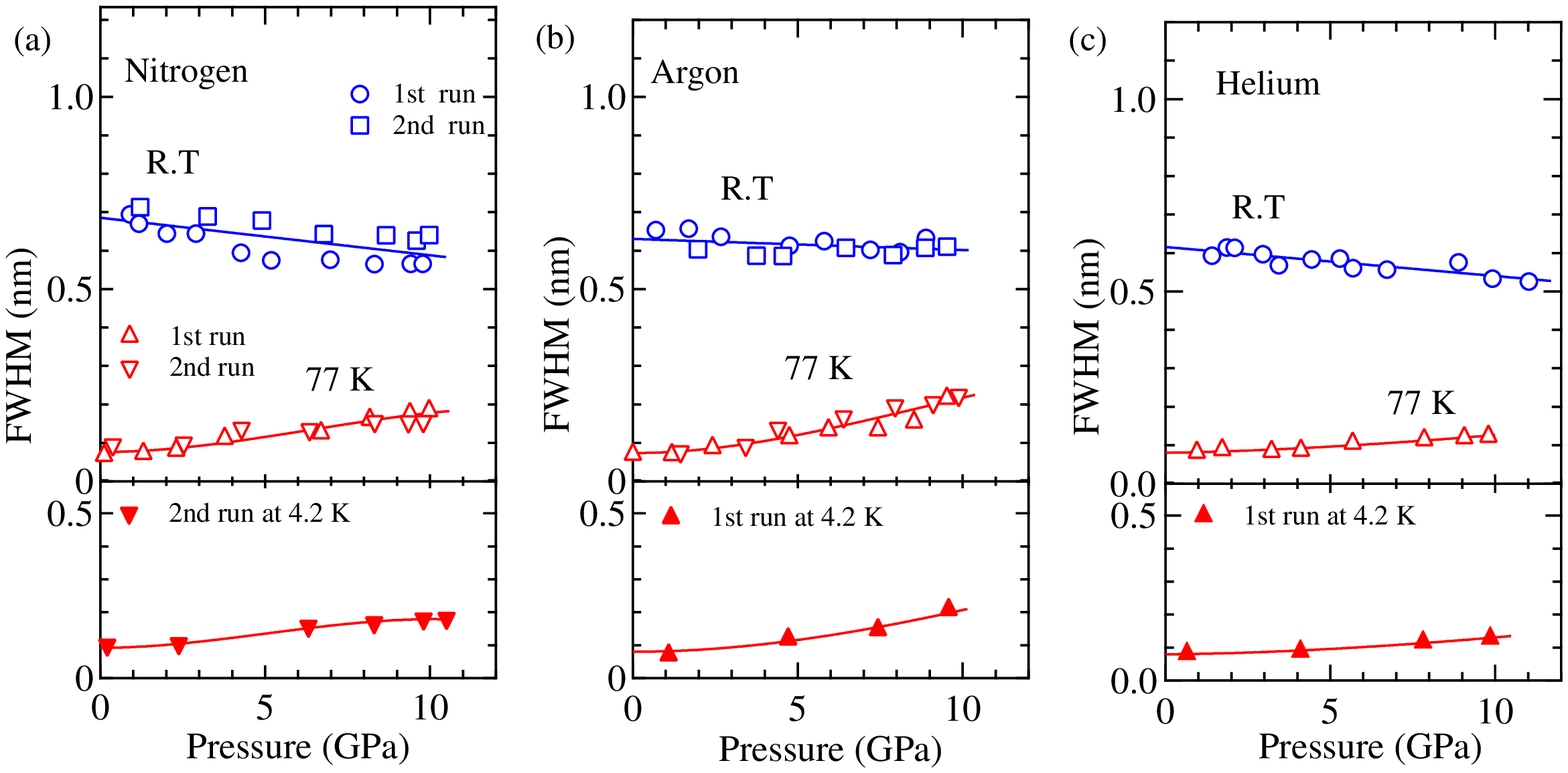}
\caption{\label{fig:epsart}(Color online) Pressure dependence of full-width at half maximum (FWHM) of the ruby $R_1$ luminescence line at room temperature and 77 and 4.2 K for (a) nitrogen, (b) argon, and (c) helium. }
\end{figure*}

 At room temperature, the FWHMs with the three media do not show any strong increase. The FWHM for the pentane mixture starts to weakly increase above the hydrostatic pressure of 7.3 GPa. The FWHM for the 4:1 M-E mixture starts to weakly increase above 8 GPa, smaller than the hydrostatic-limit pressure of 10.5 GPa determined at 297 K~\cite{piermarini1}.  It has been  reported that the liquid-glass transition pressure $P_{glass}$ is sensitive to temperature~\cite{grocholskia}. The value of $P_{glass}$ in the present experiments at $T$ = 283 ${\pm}$ 3 K is  estimated to be 8.5 ${\pm}$ 0.5 GPa, which is consistent with the present result. The FWHM for petroleum ether does not show any clear increase around $P_{glass}$ of 6 GPa~\cite{barnett2}. It is suggested that the mechanical strength (yield and shear strength) of solid petroleum ether is not strong enough to distort the ruby crystal.  At 77 K, the FWHMs for the three media weakly increase with increasing pressure below 5 GPa. The values of the $\Delta$FWHMs at 5 GPa are less than 0.1 nm. The FWHM for the pentane mixture shows a comparably stronger increase above 8 GPa. The values of the $\Delta$FWHMs for the media are less than 0.3 nm at 10 GPa.

   Finally, we show the results of nitrogen, argon and helium. It is well-known that these elements produce good hydrostatic pressure at room temperature~\cite{bell,liu,lesar,takemura4}. The hydrostatic limit pressures at room temperature were determined to be 13, 9 and 30 GPa for nitrogen, argon and helium, respectively. The previous work carried out up to 3 GPa revealed that helium does not provide any nonhydrostaticity even at 4.2 K and that argon provides slightly less hydrostaticity but with a slight improvement over the 4:1 mixture of methanol and ethanol~\cite{burnett}. 
    
  Figures 6 (a), (b) and (c) show the pressure dependences of FWHMs for nitrogen, argon and helium, respectively, at room temperature, 77 and 4.2 K. At room temperature, the FWHMs for the media weakly decrease with increasing pressure, which is a characteristic feature when the pressure is hydrostatic~\cite{piermarini1}.  At 77 K, the FWHMs for the media very weakly increase with increasing pressure. The value of the $\Delta$FWHMs for the helium medium in particular is less than 0.1 nm below 10 GPa. The values of FWHMs for the media at 4.2 K are the same as those at 77 K within experimental errors. 
    
    The present work reveals the suitability of nitrogen, argon and helium as pressure-media for cryogenic experiments with DACs. The helium medium has been frequently used in X-ray diffraction experiments. But the application of the medium in electrical transport measurement has not remarkably advanced, apart from in a few studies~\cite{hikita,thomasson1}. This is due to the large compressibility of helium in the low-pressure region~\cite{loubeyre}. Meanwhile, the compressibility of nitrogen and argon are smaller than that of helium\cite{errandonea2,gregoryanz}. The two media can be safely used in electrical transport measurements with DACs~\cite{knebel,tateiwa}. 

  It is interesting to introduce a recent work by Klotz {\it et al.} in which a weak deviation from the homogenous pressure in the sample space of the DAC was thoroughly studied using eleven kinds of pressure-media at room temperature\cite{klotz}.  They pointed out that the pressure quality of the argon medium is worse than that of the nitrogen medium above 2 GPa, even at room temperature. Unfortunately, the small pressure gradient could not be detected in this study as the broadening effect reflects the spatially averaged information on the nonhydrostatic effects in the sample chamber. It is noted, however, that the FWHM (or $\Delta$FWHM) with the argon medium is slightly larger than with nitrogen above 8 GPa at 77 K.
 \begin{figure*}
\begin{center}
\includegraphics[width=15cm]{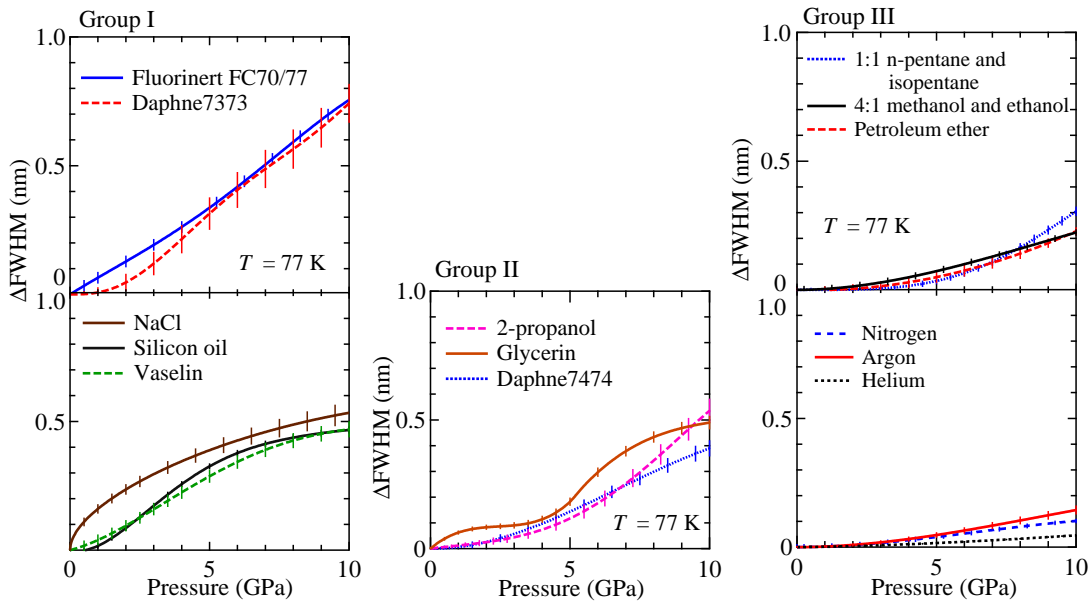}
\end{center}
\caption{\label{fig:epsart} (Color online)Pressure dependence of $\Delta$FWHM (= FWHM($P$)-FWHM(0)) of the ruby $R_1$ line for fourteen pressure-media at 77 K. Here, FWHM($P$) and FWHM(0) are the values of the fitted function for the FWHM under high pressure and ambient pressure, respectively. Error bars indicate averaged deviations of the experimental data of the FWHM from the fitted line at each pressure. }
\end{figure*} 
      
   The pressure change during the cooling process (${\Delta}P/P$ = $(P({\rm 77 K})-P({\rm R.T}))/P({\rm R.T})$) is in the order of few \% for the organic and argon media with the present DAC. However, the values of ${\Delta}P/P$ for the helium and nitrogen media are about  30 and 40 \%, respectively, at 2 GPa. The pressure change decreases with increasing pressure.  The change is in the order of a few \% above 5 GPa with the nitrogen medium and is 10 \% at 10 GPa with the helium medium. This pressure change could be a disadvantage when the temperature dependence of a physical quantity is measured in a wide temperature region. Also, a large change of pressure in the low temperature region could cause the nonhydrostatic pressure, making a damage to a sample even if the helium medium is used~\cite{goree}.

     \section{Discussion and conclusion}
In this section, the relative strength of the nonhydrostatic effects in the media is discussed from the line widths of the ruby $R_1$ line. For the sake of simplicity, the individual data points are omitted and only the guide (fitted) lines for the FWHMs at 77 K in Figures 2 -  6 are compared. To discuss the pressure effect clearly, the ambient pressure width FWHM(0) is subtracted from the guide line: $\Delta$FWHM = FWHM($P$) - FWHM(0). Here FWHM(0) was determined for each of the media. Figure 7 shows the pressure dependences of $\Delta$FWHMs for all the media. 

  The $\Delta$FWHMs for the media in Group I show the large value of above 0.3 nm at 5.0 GPa, indicating the strong nonhydrostatic effects of the media. Although some of the media in the group have advantages for use in high-pressure experiments, they are not suitable media in cryogenic experiments with the DAC at the higher-pressure region. The nonhydrostatic effects of the media in Group II are smaller than those of the media in Group I below 5 GPa. However, the effects become stronger above the pressure and are comparable with those of the media in the former group around 10 GPa.  The effects of the media in Group III are smaller than those of the media in the other groups, even in the low temperature region, up to 10 GPa. In particular, the broadening effect of the ruby $R_1$ line in the nitrogen, argon and helium media are significantly small, suggesting that the media are more appropriate for cryogenic experiments with the DAC.

 Finally, we mention some notes for interpretation of the present results from different points of views.
   
  (1) The present results are applicable in high-pressure studies using opposed anvil type high-pressure cells (DAC and Bridgman cells) but not simply in studies using multi-anvil type high-pressure cells such as a cubic anvil cell or Kawai-type (6-8) multi anvil high-pressure cell where a sample with a medium is pressed by six or eight anvils\cite{mori2,fujioka,kawai2}. Pressurizing a sample highly symmetrically generally reduces the nonhydrostaticity of the pressure caused by a solid pressure medium and therefore achieves a good hydrostatic pressure. 
      
  (2) Ruby crystals are relatively stiff material with a bulk modulus of 253 GPa~\cite{syassen}. If the material to be studied is strain sensitive or has mechanical strength smaller than the ruby crystal, the nonhydrostatic effects could influence the physical properties of the material, even though the broadening of the ruby $R_1$ line would be small or absent. There are usually discrepancies between hydrostatic-limit pressures determined using different methods for one pressure-medium. An extreme example is the nonhydrostatic effects on elastically soft quartz (SiO$_2$) with a bulk modulus of 37.1 GPa~\cite{angel}. 
    
  (3) Some of the organic media in Groups II and III could chemically react with organic materials. There is a possibility that small argon and helium atoms could penetrate into samples with a cage structure, like fullerenes of C$_{60}$ and C$_{70}$~\cite{samara}.

  \section{Acknowledgments}
This work was financially supported by a Grant-in-Aid for Scientific Research on Innovative Areas "Heavy Electrons" (No. 20102002), Scientific Research of Priority Area, Creative Scientific Research (15GS0213) and Scientific Research (S and B) made by the Ministry of Education, Culture, Sports, Science and Technology (MEXT) and Japan Society of the Promotion of Science (JSPS).

\bibliography{apssamp}

\begin{references}

\bibitem{jayaraman1} A. Jayaraman, Rev. Mod. Phys. {\bf 55}, 65 (1983). 
\bibitem{eremets}  M. I. Eremets,  {\it High Pressure Experimental Methods} (Oxford University Press, Oxford, 1996). 
\bibitem{buzea} C. Buzea and K. Robble, Supercond. Sci. Technol. {\bf 18}, R1 (2005). 
\bibitem{shimizu1} K. Shimizu, J. Phys.: Condens. Matter  {\bf 19}, 125207 (2007). 

\bibitem{flouquet} J. Flouquet,  {\it Progress in Low Temperature Physics,} ed by W. P. Halperin (Elsevier, Amsterdam, 2005), Vol. {\bf 15}, Chap. 2.
\bibitem{onuki} {\it Frontiers of Novel Superconductivity in Heavy Fermion Compounds}, Special Topics of J. Phys. Soc. Jpn. {\bf 76}, No.5 edited by Y. {\=O}nuki and Y. Kitaoka (Physical Society of Japan, Tokyo, 2007) 

\bibitem{torikachvili}M. S. Torikachvili, S. L. Bud'ko, N. Ni, and P. C. Canfield, Phys. Rev. Lett. {\bf 101}, 057006 (2008).
\bibitem{park}T. Park, E. Park, H. Lee, T. Klimczuk, E. D. Bauer, F. Ronning, and J. D. Thompson, J. Phys.: Condens. Matter  {\bf 20}, 322204 (2008).
\bibitem{yu} W. Yu, A. A. Aczel, T. J. Williams, S. L. Bud'ko, N. Ni, P. C. Canfield, and G. M. Luke, Phys. Rev. B {\bf 79}, 020511(R) (2009).

\bibitem{piermarini1}G. J. Piermarini, S. Block, and J. D. Barnett, J. Appl. Phys. {\bf 44}, 5377 (1973). 
\bibitem{angel}R. J. Angel, M. Bujak, J. Zhao, G. D. Gatta, and S. D. Jacobsen, J. Appl. Crystallogr. {\bf 40}, 26 (2007).
\bibitem{osakabe}T. Osakabe and K. Kakurai, Jpn. J. Appl. Phys. {\bf 47}, 6544 (2008).
\bibitem{klotz}S. Klotz, J. C. Chervin, P. Munsch, and G. Le Marchand, J. Phys. D: Appl. Phys. {\bf 42}, 075413 (2009).
\bibitem{burnett}J. H. Burnett, H. M. Cheong, and W. Paul, Rev. Sci. Instrum.  {\bf 61}, 3904 (1990). 
\bibitem{fukazawa}H. Fukazawa, K. Hirayama, T. Yamazaki, Y. Kohori, and T. Matsumoto, J. Phys. Soc. Jpn.  {\bf 76}, 125001 (2007). 
\bibitem{syassen}K. Syassen, High Pressure Research {\bf 28}, 75 (2008).
\bibitem{dunstan1} D. J. Dunstan and I. L. Spain, J. Phys. E: Sci. Instrum. {\bf 22}, 913 (1989). 
\bibitem{dunstan2} I. L. Spain and D. J. Dunstan, J. Phys. E: Sci. Instrum. {\bf 22}, 923 (1989). 
\bibitem{koyama} K. Koyama-Nakazawa, M. Koeda, M. Hedo, and Y. Uwatoko, Rev. Sci. Instrum.  {\bf 78}, 066109 (2007). 

\bibitem{chai}M. Chai and J. Michael Brown, Geophys. Res. Lett. {\bf 23}, 3539 (1996).
\bibitem{adams}D. M. Adams, R. Appleby, and S. K. Sharma, J. Phys. E: Sci. Instrum. {\bf 9}, 1140 (1976).
\bibitem{he}J. He and D. R. Clarke, J. Am. Ceram. Soc. {\bf 78}, 1347 (1995).

\bibitem{zha}C. S. Zha, H. K. Mao, and R. J. Hemley, Proc. Natl Acad. Sci. USA {\bf 97}, 13494 (2000).
\bibitem{kawamura}K. Nakao, Y. Akahama, Y. Ohishi, and H. Kawamura, Jpn. J. Appl. Phys. {\bf 39}, 1249 (2000).
 
\bibitem{sidorov1}V. A. Sidorov and R. A. Sadykoy, J. Phys.: Condens. Matter {\bf 17}, S3005 (2005). 
\bibitem{murata1}K. Murata, H. Yoshino, H.O.Yadav, Y. Honda, and N. Shirakawa, Rev. Sci. Instrum.  {\bf 68}, 2490 (1997). 
\bibitem{murata2}K. Yokogawa, K. Murata, H. Yoshino, and S. Aoyama, Jpn. J. Appl. Phys. {\bf 46}, 3636 (2007). 

\bibitem{bridgman1}P. W. Bridgman, Proc. Am. Acad. Arts Sci.  {\bf 77}, 129 (1949). 
\bibitem{kirichenko}A. S. Kirichenko, A. V. Kornilov, and V. M. Pudalov, Instrum. and Exp. Tech.  {\bf 48}, 813 (2005). 
\bibitem{sandberg}O. Sandberg and B. Sundqvist, J. Appl. Phys.  {\bf 53}, 8751 (1982). 

\bibitem{wittig1}N. Lotter and J. Wittig, J. Phys. E: Sci. Instrum. {\bf 22}, 440 (1989). 
\bibitem{sacchetti}A. Sacchetti, M. C. Guidi E. Arcangeletti, A. Nucara, P. Calvani, M. Piccinini, A. Marcelli, and P. Postorino, Phys. Rev. Lett. {\bf 96}, 035503 (2006).

\bibitem{ragan}D. D. Ragan, D. R. Clarke, and D. Schiferl, Rev. Sci. Instrum.  {\bf 67}, 494 (1996). 
\bibitem{shen1}Y. Shen, R. S. Kumar, M. Pravica, and M. F. Nicol, Rev. Sci. Instrum.  {\bf 75}, 4450 (2004). 

\bibitem{bridgman0}P. W. Bridgman, Rev. Mod. Phys. {\bf 18}, 1 (1946).
\bibitem{kurita1} N. Kurita, M. Hedo, M. Kato, T. Fujiwara, Y. Uwatoko, and S. R. Tozer, J. Magn. Magn. Mater {\bf 310}, 611(2007). 
\bibitem{murata3}K. Murata, K. Yokogawa, H. Yoshino, S. Klotz, P. Munsch, A, Irizawa, M. Nishiyama, K. Iizuka, T. Nanba, T. Okada, Y. Shirage, and S. Aoyama, Rev. Sci. Instrum.  {\bf 79}, 085101 (2008). 

\bibitem{drozd}A. Drozd-Rzoska, S. J. Rzoska, M. Paluch, A. R. Imre, and C. M. Roland,  J. Chem. Phys. {\bf 126}, 164504 (2007).

\bibitem{bridgman2}P. W. Bridgman, Proc. Am. Acad. Arts Sci.  {\bf 77}, 117 (1949). 

\bibitem{bancroft}D. Bancroft, Phys. Rev. {\bf 53} 587 (1938). 

\bibitem{sawaoka}A. Sawaoka and N. Kawai, Jpn. J. Appl. Phys.  {\bf 9}, 353 (1970). 


\bibitem{grocholskia}B. Grocholski and R. Jeanloz, J. Chem. Phys. {\bf 123}, 204503 (2005).
\bibitem{barnett2}J. D. Barnett and C. D. Bosco, J. Appl. Phys. {\bf 40}, 3144 (1969). 



\bibitem{bell}P.  M. Bell and H. K. Mao, Carnegie Inst. Wash. Year book. {\bf 80}, 404 (1982).
\bibitem{liu}Z. Liu, Q. Cui, and G. Zou, Physics Letters A {\bf 143}, 79 (1990).
\bibitem{lesar}R. LeSar, S. A. Ekberg, L. H. Jones, R. L. Mills, L. A. Schwalbe, and D. Schifer  Solid State Commun. {\bf 32}, 131 (1979).
\bibitem{takemura4}K. Takemura and A. Dewaele, Phys. Rev. B  {\bf 78}, 104119 (2008). 

\bibitem{hikita}T. Hikita, T. Maruyama, and N. Yamada, Jpn. J. Appl. Phys. {\bf 29}, 2519 (1990). 
\bibitem{thomasson1} J. Thomasson, Y. Dumont, J.-C. Griveau, and C. Ayache, Rev. Sci. Instrum.  {\bf 68}, 1514 (1997). 
\bibitem{loubeyre}P. Loubeyre, R. LeToullec, J. P. Pinceaux, H. K. Mao, J. Hu, and R. J. Hemley, Phys. Rev. Lett.  {\bf 71}, 2272 (1993). 

\bibitem{errandonea2}D. Errandonea, R. Boehler, S. Japel, M. Mezouar, and L. R. Benedetti, Phys. Rev. B. {\bf 73}, 092106 (2006).

\bibitem{gregoryanz}E. Gregoryanz, A. F. Goncharov, C. Sanloup, M. Somayazulu, H-k. Mao, and R. J. Hemley, J. Chem. Phys.  {\bf 126}, 184505 (2007). 
\bibitem{knebel}G. Knebel, M. A, M\'{e}asson, B. Salce, D. Aoki, D. Braithwaite, J. P, Brison, and J. Flouquet, J. Phys.: Condens. Matter {\bf 16}, 8905 (2004). 
\bibitem{tateiwa}N. Tateiwa, S. Ikeda, Y. Haga, T. D. Matsuda, M. Nakashima, D. Aoki, R. Settai, and Y. {\=O}nuki, Journal of Physics: Conference Series {\bf 150}, 042206 (2009).  

\bibitem{goree}W. S. Goree and T. A. Scott, J. Phys. Chem. Solids {\bf 27}, 835 (1966). 


\bibitem{mori2}N. Mori, H. Takahashi, and N. Takeshita, High Pressure Research {\bf 24}, 225 (2004). 
\bibitem{fujioka} N. Fujioka, O. Mishima, W. Endo, and N. Kawai, J. Appl. Phys. {\bf 49} 4830 (1978). 
\bibitem{kawai2} N. Kawai, H. Sakamoto, Y. Notsu, and A. Onodera, Proc. Japan. Acad. {\bf 51}, 623 (1975). 


\bibitem{samara}G. A. Samara, L. V. Hansen, R. A. Assink, B. Morosin, J. E. Schirber and D. Loy, Phys. Rev. B {\bf 47}, 4756 (1993).


\end{references}


\end{document}